\documentclass[apjl]{emulateapj}

\usepackage{url}
\usepackage{color}
\usepackage{ulem}

\shorttitle{Starburst in the Circumnuclear Disk of NGC~1275}
\shortauthors{Nagai \& Kawakatu}

\begin{document}


\title{Diffuse Synchrotron Emission Associated with the Starburst in the Circumnuclear Disk of NGC~1275}


\author{H. Nagai\altaffilmark{1,2} 
\& N. Kawakatu\altaffilmark{3}, 
}



\altaffiltext{1}{National Astronomical Observatory of Japan, Osawa 2-21-1, Mitaka, Tokyo 181-8588, Japan}\email{hiroshi.nagai@nao.ac.jp}
\altaffiltext{2}{The Graduate University for Advanced Studies, SOKENDAI, Osawa 2-21-1, Mitaka, Tokyo 181-8588, Japan}
\altaffiltext{3}{National Institute of Technology, Kure College, 2-2-11, Agaminami, Kure, Hiroshima, 737-8506, Japan}
%

\begin{abstract}
Recent Atacama Large Millimeter/submillimeter Array (ALMA) observations found a positive correlation between the mass of dense molecular gas in the circumnuclear disks (CNDs) and accretion rate to the active galactic nuclei (AGNs).  This indicates that star formation activity in the CNDs is essential for triggering the accretion of mass to AGNs.  Although the starburst-driven turbulence is a key mechanism for the transfer of angular momentum and the resultant mass accretion from the CND scale to the inner radius, the observational evidence is lacking.  We report the very-long-baseline-interferometry (VLBI) detection of the diffuse synchrotron emission on a scale of several tens-pc coinciding spatially with the molecular gas disk recently discovered by ALMA observations in NGC~1275.  The synchrotron emissions are most likely resulted from the relativistic electrons produced by the supernova explosions.  This is an unambiguous evidence of the star formation activity in a CND.  The turbulent velocity and the scale height of the CND predicted from the supernova-driven turbulence model agree with the observations, although the model-predicted accretion rate disagrees with the bolometric luminosity.  This might indicate that additional mechanisms to enhance the turbulence is required for the inner disk.  We discuss the multiphase nature of the CND by combining the information of the CO emission, synchrotron emission, and free-free absorption.
\end{abstract}		


\keywords{galaxies: active, galaxies: nuclei, galaxies: elliptical and lenticular, cD, galaxies: individual (3C~84, NGC~1275, Perseus~A), radio continuum: galaxies, radio lines: galaxies 
}	
\section{Introduction}\label{sect:intro}
Recent Atacama Large Millimeter/submillimeter Array (ALMA) observations have revealed the presence of massive molecular gas around active galactic nuclei (AGNs).  Dozens of sources show a disk-like structure (i.e., so-called circumnuclear disk (CND)) decoupled, both morphologically and kinematically, from the extended galactic-scale molecular gas  \citep[e.g.][]{Imanishi2018, Izumi2018, Combes2019}.  There is a positive correlation between the mass of dense molecular gas in CNDs and AGN accretion rates \citep{Izumi2016}.  This suggests that these molecular CNDs are important components to understand the accretion physics in AGNs.  

The angular momentum transfer mechanism that enhances the mass accretion from the CND scale to the inner radius has been a subject of discussion.  One plausible explanation is that gas turbulence generated by supernova (SN) explosions in the CNDs/tori (i.e., so-called SN-driven turbulence model) has a major role for the transfer of angular momentum \citep{Kawakatu2008}.  However, there has been a lack of direct observational evidence of the star formation activity in the CNDs/tori.  Polycyclic aromatic hydrocarbons (PAHs) are widely used as tracers of star formation \citep[e.g.,][]{Peeters2004}, but PAHs in the CNDs/tori can be dissociated by harsh AGN radiation.  Instead, synchrotron radiation from relativistic electrons produced by SN explosions can be an alternative tracer.  For this to be possible, spatially-resolved imaging is critical to study the spatial association of the synchrotron emission with molecular CND/tori and decouple the emission from the AGN jets.

NGC~1275 (z=0.01755) is a nearby radio galaxy/giant elliptical galaxy at the center of the Perseus cluster.  The radio source of NGC~1275 is also known as 3C~84 and shows multiple radio lobes on different angular scales \citep{Silver1998, Walker2000, Nagai2010}.  This galaxy is a reservoir of a large amount of cold gas \citep[$M_{\rm gas}\sim10^{9}~M_{\sun}$:][]{Lim2008, Salome2008}.  On kpc scales, molecular gas filaments are aligned in the east-west direction \citep{Lim2008} and these filaments seem to coincide with the H$\alpha$ nebulae \citep{Fabian2008}.  Within 100~pc, recent ALMA observations detected a rotating CND with the CO(2-1), HCN(3-2), and HCO$^{+}$(3-2) lines \citep{Nagai2019}.  Intriguingly, the detection of diffuse synchrotron emission on this spatial scale was previously reported by very long baseline interferometry (VLBI) observations at the frequencies $\lesssim1$~GHz \citep{Silver1998}.  Both relativistic electrons diffused out from the AGN and those produced by the SN explosions were discussed as the origin of the diffuse synchrotron emission, but the actual origin was unclear because of a lack of the molecular CND information at that time.  

Here we report the detailed comparison of the molecular CND with the diffuse synchrotron emission.  Throughout this paper, we used $H_{0}$=69.6, $\Omega_{\rm M}=0.286$, and $\Omega_{\Lambda}=0.714$.  At the of NGC~1275 distance, $0.1\arcsec$ corresponds to 34.4~pc.

\section{OBSERVATIONS and DATA ANALYSIS}\label{sect:Obs}
\subsection{The ALMA Data}
The observations were done in ALMA Band 6 ($\lambda=$1.3~mm) to cover the CO(2-1) line with 47 antennas on 2017 November 27.  The maximum and minimum baseline lengths were 8.5~km and 92.1~m, respectively.  The observations consisted of 52 scans with a 54.4-sec integration for each scan bracketed by the scans for the observations of the complex gain calibrator.  Integration time per an interferometric visibility was set to 2 sec. The data were processed at East Asia ALMA Regional Center (EA-ARC) with a standard manner using the software CASA 5.1.1-5 and ALMA Pipeline version 40896.  Images were created with a velocity resolution of 20~km/s with the ``nearest" interpolation in frequency.  Deconvolution was performed with the CLEAN algorithm using the CASA task \texttt{tclean} non-interactively.  We performed self-calibration using the continuum emission of NGC~1275 in both phase and amplitude that improved the image quality significantly. The phase and amplitude self-calibration were done per integration time and per scan, respectively.  Final image rms is 0.87~mJy~beam$^{-1}$ with natural weighting.  The beam size is ($144\times77$)~mas at a position angle of $4.0\degr$.  Further details on the observations and data analysis are provided in \cite{Nagai2019}.
 
\subsection{The VLBA Data}
\begin{figure*}
\begin{center}
\includegraphics[width=15cm]{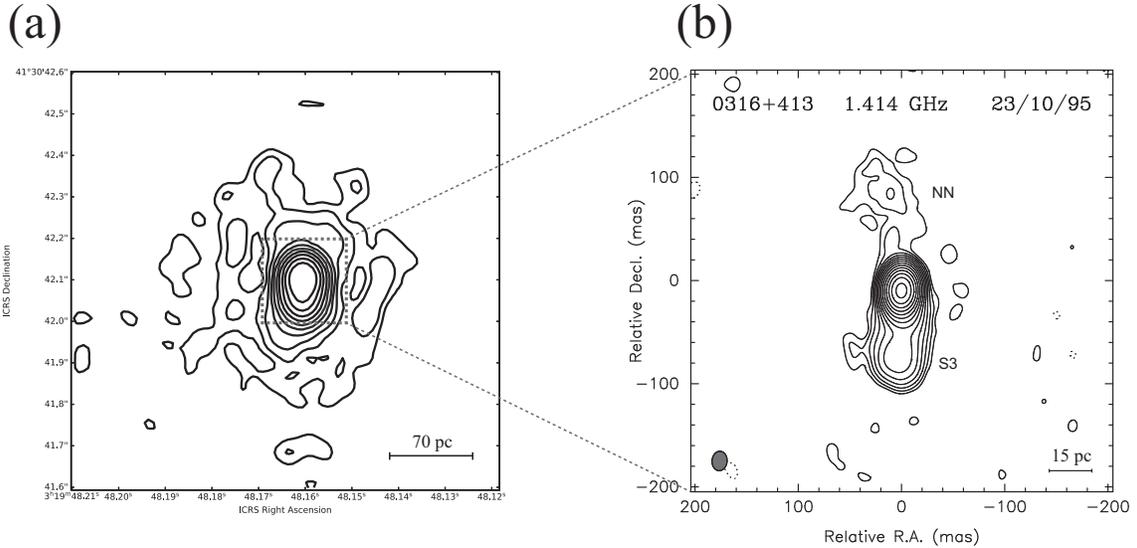}
\end{center}
\caption{(a) Contours of 330-MHz continuum image.  The restoring beam is $(46.6\times39.8)$~mas at $4.05\degr$.  The peak intensity is 2.5~Jy~beam$^{-1}$.  The contours are plotted at the level of $3\sigma\times(-1, 1, 2, 4, 8, 16, 32, 64, 128, 256, 512)$, where $\sigma$ is the image noise RMS of 0.573~mJy~beam$^{-1}$. The gray broken square indicates the area of Fig. \ref{fig:fig1}(b). (b) Contours of 1.4~GHz continuum image \citep{Silver1998} that highlights the inner jet structure (\copyright AAS. Reproduced with permission).}
\label{fig:fig1}
\end{figure*}

We analyzed an archival Very Long Baseline Array (VLBA) data.  The observation was carried out at 330~MHz with the 10 VLBA stations on 1995 October 23.  The net integration time for NGC~1275 was 250 minutes.  Standard data flagging, bandpass and amplitude calibrations, and fringe fitting were performed with the Astronomical Image Processing System (AIPS).  Image deconvolution and self-calibration were performed using the software DIFMAP.  We began self-calibration with a long solution interval and subsequently shortened the solution interval down to 1 min.  Figure \ref{fig:fig1} shows the CLEANed image.  The image rms is 0.573~mJy~beam$^{-1}$ with natural weighting.  The beam size is ($47\times40$)~mas at $4\degr$.  The image was previously reported in \cite{Silver1998} and we reproduced a consistent image in terms of its overall structure, the image noise rms, and peak intensity. 

\section{RESULTS}\label{sect:results}
\begin{figure}
\begin{center}
\includegraphics[width=8.5cm]{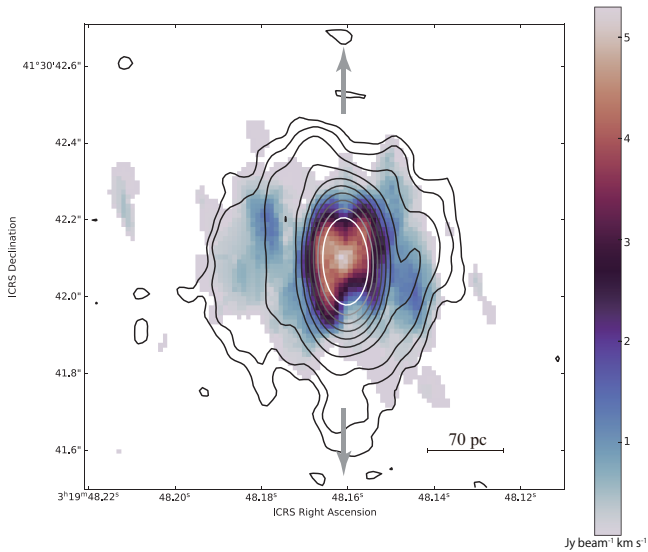}
\end{center}
\caption{Contour map at 330-MHz intensity overlaid on the velocity integrated intensity (moment 0 map) of the CO(2-1) in color.  The restoring beam of the CO(2-1) moment 0 map is $(144\times77)~{\rm mas}$ at $4.0\degr$.  The images are plotted throughout the region where the flux density of the image cube with a velocity resolution of 20~km~s$^{-1}$ is greater than 5$\sigma$ ($1\sigma=0.87$~mJy).  The contour map is also convolved with the same restoring beam.  The gray arrows in the north and south indicate an approximate direction of the jet shown in Fig. \ref{fig:fig1}(b).}
\label{fig:fig2}
\end{figure}

Figure \ref{fig:fig2} shows the 330-MHz continuum map overlaid on the velocity integrated map (moment 0) of the CO(2-1).  The velocity integrated image of the CO(2-1) is the same image presented in \cite{Nagai2019}.  The image was created with natural weighting.  The image shows a disk-like morphology with a radius of $\sim100$~pc.  There are clumpy substructures within the disk that suggest its nonuniformity.  \cite{Nagai2019} reported the velocity structure of the disk and found a velocity gradient along the position angle of $\sim70\degr$.  This velocity gradient is most likely explained by the disk rotation.  A similar morphology and velocity structure are also seen in the images of HCN(3-2) and HCO$^{+}$.  More details about the properties of the molecular CND can be found in \cite{Nagai2019}.

The continuum image at 330~MHz shows the diffuse emission (denoted ``millihalo" in \cite{Silver1998}) surrounding the central bright component.  It has been reported that the central component appears to be dominated by the jet/lobe emission with a projected length of 30~pc at higher frequencies \citep{Silver1998, Romney1995}.  The flux density of the diffuse emission is mostly below 10~mJy~beam$^{-1}$ ($\sim0.5$\% of the peak intensity).  \cite{Silver1998} reported a power-law spectrum for the diffuse continuum emission, which demonstrates that it has a nonthermal origin. Remarkably, the diffuse continuum emission shows a good spatial coincidence with the molecular CND at 30-100~pc.

Absolute position information was lost due to the self-calibration for both the VLBA and ALMA images; thus, the center of the VLBA image may not be the same as that of the ALMA image.  The center of the VLBA image must be the position where the 330-MHz emission is the strongest.  This position probably coincides with S1 \citep{Silver1998}, which is offset from the VLBI core by $\sim10$~mas.  The center of the ALMA image must conincide with the VLBI core where the millimeter continuum emission is expected to be the strongest \citep[See discussion in][]{Nagai2017}.  This could result in an positional mismatch between the VLBA and ALMA images at a level of 10~mas.  However, this level of offset is smaller than the size of restoring beam.  The spatial coincidence of the molecular CND with the 330-MHz synchrotron emission should be still watertight.

\section{DISCUSSION}
\subsection{Origin of the Diffuse Synchrotron Emission}
The spatial coincidence of molecular CND with diffuse continuum emission indicates that both molecular gas and relativistic electrons are co-located in the same disk.  What is the origin of the relativistic electrons?  One possible answer is the relativistic electrons that are accelerated by the SN explosion, while another is those provided by the AGN. 

The AGN of NGC~1275 shows a prominent radio-jet activity \citep{Nagai2010, Nagai2014} that can supply the relativistic electrons.  For the Fanaroff-Riley Type I (FRI) AGNs \citep{Fanaroff1974}, jets gradually loosen their collimation as they propagate;  as a result, the jets become wider at their downstream, but the overall structure generally maintains the two-sided shape that does not resemble the observed diffuse radio emission in NGC~1275.  For the Fanaroff-Riley Type II (FRII) AGNs \citep{Fanaroff1974}, the jets are well collimated up to the termination points (hotspots), and the shocked gas (relativistic electrons) leak out therefrom. The sideway-expansion of the shocked gas and propagation of the hotspots can make the overall structure ``cocoon" shape \citep{Begelman1989}.  One might think that the observed diffuse radio emission in NGC~1275 can be explained the the cocoon emission.  However, the emission from the cocoon must be brighter on the both sides of hotspot and dimmer near the AGN because of the cooling of electrons by synchrotron radiation \citep[e.g.][]{Carilli1991}.  Furthermore, many of FRII sources show that the cocoon is longer in the jet direction \citep{Alexander1987} because the propagation speed of hotspots is faster than the speed of sideway-expansion.  Therefore, it seems to be difficult to explain the relatively symmetric shape and uniform brightness of the diffuse radio emission in NGC~1275 as shocked gas in the cocoon associated with the jets on tens pc scale (shown in Figure \ref{fig:fig1}(b)) or jets extending further out \citep{Pedlar1990}.  In summary, the relativistic electrons carried by AGN jets cannot be the reason for the diffuse radio emission in NGC~1275.

Relativistic electrons that is produced near the jet base may be transported outward from the AGN.  The transported electrons could propagate preferentially over the disk if the toroidal magnetic field was predominant in the disk.  \cite{Silver1998} derived an Alfven velocity of $\sim430$~km~s$^{-1}$ using an equipartition magnetic field of 200~$\mu$G and gas density of 1~cm$^{-3}$.  This gas density was derived using equipartition condition between the relativistic plasma and the thermal gas pressures at a gas temperature of $\sim10^{7}$~K \citep{Fabian1981}.   It was concluded that the derived velocity was close to that required to supply relativistic electrons to the edge of the source within the electron cooling timescale of $3.2\times10^{5}$~yr at 330~MHz without requiring reacceleration.  This can be true if only the hot plasma radiating X-ray emission at the core of the Perseus cluster is responsible for the gas pressure.  
However, the gas density should be much higher than the number used for the calculation by \cite{Silver1998} because of the presence of the molecular CND with a gas density of $10^{3}$-$10^{4}$~cm$^{-3}$ \citep{Scharwachter2013, Nagai2019}.  Thus, the electrons cannot reach the edge of the CND.  Therefore, the diffuse radio emission cannot originate from the AGN.  The most likely origin is that the relativistic electrons accelerated by the SN explosions in the CND.

\cite{Silver1998} reported that the radio luminosity of the CND shows a good agreement with the infrared luminosity as seen in normal galaxies \citep{Condon1992}.  The inferred SFR is 3~$M_{\sun}$~yr$^{-1}$ \citep{Silver1998} (Note that \cite{Silver1998} derived  SFR=6.4~$h^{-2}~M_{\sun}$~yr$^{-1}$ where $H_{0}=100h$~km~s$^{-1}$~Mpc$^{-1}$).  We note that this SFR is consistent with that constrained by the observation of PAHs and far-infrared luminosity \citep[SFR$<8.22~M_{\sun}$~yr$^{-1}$:][]{Oi2010}.  For the case of Milky way galaxy, number of different observations suggest that the SN rate is about a hundredth of the SF rate  \citep[see Table 1 in][]{Diehl2006}.  If we apply this relation to the NGC~1275, $\sim10^4$ of SNs should have happened within the electron cooling timescale.  Higher angular resolution observations might be able to resolve individual SNs, as demonstrated in NGC~7469 \citep{Colina2001} and Arp~220 \citep{Lonsdale2006}.  It would be also important to resolve the CND by near-infrared continuum observations with the adoptive optics to obtain further confirmation of the SF activity in the CND.

\subsection{Comparison with the SN-driven Turbulence Model}
\cite{Kawakatu2020} provided a turbulent velocity ($v_{t}$) in the CND (see equation (8) in \cite{Kawakatu2020}) based on the SN-driven turbulence model, where the turbulence is driven by the energy input from the SN explosion.  For NGC~1275, we take a molecular gas mass ($M_{\rm gas}$) of $\sim10^{8}~M_{\sun}$\citep{Nagai2019}, star formation rate (SFR) of 3~$M_{\sun}$~yr$^{-1}$ \citep{Silver1998}, black hole mass ($M_{\rm BH}$) of $1\times10^{9}~M_{\sun}$\citep{Nagai2019}, and the disk radius ($r$) of 100~pc \citep{Nagai2019}.  Using these estimates, we obtain the following:
\begin{equation}
    v_{t}=24~{\rm km~s^{-1}}\left[\frac{C_{*}}{3\times10^{-8}~{\rm yr^{-1}}} \right]^{1/2} \left[ \frac{M_{\rm BH}}{10^{8}M_{\sun}} \right]^{-1/4} \left[ \frac{r}{{\rm 100~pc}} \right]^{3/4},
\end{equation}
where $C_{*}={\rm SFR}/M_{\rm gas}$, which agrees well with the observed velocity dispersion of $\sim25$~km~s$^{-1}$ on the $\sim100$-pc scale of the CND \citep{Nagai2019}.  This supports the idea that the SN explosion is the main driver of the turbulence in the CND.

\cite{Kawakatu2020} also gave the thickness of the CND.  For the case of NGC~1275, the equation (9) in \cite{Kawakatu2020} can be written as
\begin{equation}
    \frac{h(r_{\rm out})}{r_{\rm out}}\sim0.08 \left[\frac{C_{*}}{3\times10^{-8}~{\rm yr^{-1}}} \right]^{1/2}
\left[ \frac{M_{\rm BH}}{10^{8}M_{\sun}} \right]^{-3/4} \left[ \frac{r_{\rm out}}{{\rm 100~pc}} \right]^{5/4},
\end{equation}
where $r_{\rm out}$ is the outer radius of the CND and $h(r_{\rm out})$ is the scale height of the disk at $r=r_{\rm out}$.  Therefore, the CND of NGC~1275 is expected to be thin.  This agrees with the signature of the low-covering fraction of fluorescent FeK$\alpha$ matter detected by X-ray observations \citep{Hitomi2018}.  With $h(r=100~{\rm pc})\simeq10$~pc, the dissipation timescale of the turbulence ($t_{\rm dis}=h/v_t$, see \cite{Kawakatu2020}) is estimated to be $4\times10^6$~yr.  About $10^{5}$ SNs should have happened within this timescale if we take SFR$=3~M_{\sun}$~yr$^{-1}$ \citep{Silver1998} and SN rate of one hundredth of the SF rate \citep{Diehl2006}.  This indicates that the SN explosions must continuously supply the energy in the CND against the energy dissipation.  Therefore, the diffuse synchrotron emission should not be the fossil of old SNs but young enough to account for the CO velocity dispersion.

We can also compare the black hole accretion rate derived from the observed bolometric luminosity with the accretion rate at the inner radius of the CND ($\dot{M}_{\rm acc}(r_{\rm in})$) predicted by the SN-driven turbulence model.  Using the equation (8) in \cite{Izumi2016}, we can write $\dot{M}_{\rm acc}(r_{\rm in})$ of NGC~1275 as
\begin{eqnarray}
    \left[ \frac{\dot{M}_{\rm acc}(r_{in})}{M_{\sun}~{\rm yr}^{-1}} \right] &\sim& 3\times10^{-4} \left[ \frac{r_{\rm in}}{\rm 3~pc} \right]^{3} \left[ \frac{r_{\rm out}}{\rm 100~pc} \right]^{-2} \nonumber \\
    &&\left[ \frac{C_{*}}{3\times10^{-8} {\rm yr}^{-1}} \right] \left[ \frac{M_{\rm gas}}{M_{\rm BH}}/0.1 \right].
\end{eqnarray}
For simplicity, we have assumed that the gas is uniformly distributed over the CND.  The observed bolometric luminosity is $4\times10^{44}$~erg~s$^{-1}$, which gives an accretion rate of $6\times10^{-2}$ if the radiation efficiency of 0.1.  Thus, the model-predicted accretion rate is approximately three orders of magnitude smaller than the black hole accretion rate inferred from the bolometric luminosity  \citep[$4\times10^{44}$ erg~s$^{-1}$:][]{Levinson1995}.  This might indicate the nonuniformity of the CND that results in a strong time variation of the accretion rate or additional mechanisms to enhance the turbulence, such as a more intense starburst, for the inner disk.  It might also suggest that other channels of gas accretion, such as chaotic cold accretion \citep{Gaspari2013}, increase the black hole accretion rate.

Similar CND properties have been observed in the ultra-luminous infrared galaxy (ULIRG) Mrk~231.  Diffuse synchrotron emission around the AGN is observed in the central kpc \citep{Carilli1998, Taylor1999}, which  coincides spatially with the molecular CND \citep{Bryant1996, Downes1998, Feruglio2015}.  \cite{Taylor1999} concluded that the diffuse synchrotron emission is most probably related to the star formation activity in the CND.  Interestingly, the radio lobe/jet morphology is very similar to that of 3C~84 (the radio source of NGC~1275).  Both sources show a pair of compact radio lobes/jets where the emission from the counterjet is free-free absorbed by the disk \citep{Ulvestad1999}.  Using the equation (3), we obtain $\dot{M}_{\rm acc}(r_{\rm in})\sim6\times10^{-2}~M_{\sun}$~yr$^{-1}$ for Mrk~231 with $r_{\rm out}=200$~pc \citep{Feruglio2015}, SFR=220~$M_{\sun}$~yr$^{-1}$ \citep{Taylor1999}, $M_{\rm gas}\sim10^{9}~M_{\sun}$ \citep{Downes1998}, and $M_{\rm BH}\sim10^{8}~M_{\sun}$ \citep{Kawakatu2007, Yan2015}.  This is somewhat smaller than the black hole accretion rate ($\sim1.5~M_{\sun}$~yr$^{-1}$) inferred from the bolometric luminosity of $10^{46}$~erg~s$^{-1}$ \citep{Veilleux2009}.  Probing the star formation activity at innermost CND region would be important to obtain a better estimate of the accretion rate for both NGC~1275 and Mrk~231 in the future.

\subsection{The CND Structure}
The emission from the northern jet (receding jet) should pass through the CND if the CND plane is perpendicular to the jet axis\footnote{There are several estimates of the viewing angle of the jet ($\theta{\rm jet}$).  Here we assume that $\theta_{\rm jet}=65\degr\pm16\degr$ \citep{Fujita2017}.}.  \cite{Silver1998} found that the integrated spectrum of the diffuse synchrotron emission from 330~MHz to 1.4~GHz is fairly straight in contrast to the curved spectrum of the northern jet component (labeled NN), which is located at a projected distance of $\sim35$~pc ($\sim0.1\arcsec$) from the core.  The curved spectrum likely indicates the free-free absorption (FFA).  This difference in the spectral shape can provide insights into the structure of the CND.  The free-free absorption opacity can be approximated as
\begin{equation}
    \tau\approx 0.34\biggl{(}\frac{T}{10^{4}\mathrm{~{}K}}
\biggr{)}^{-1.35}\biggl{(}\frac{\nu}{\mathrm{330~MHz}}\biggr{)}^{-2.1}\biggl{(}
\frac{\mathrm{EM}}{10^{4}~\mathrm{pc~{}cm}^{-6}}\biggr{)},
\end{equation}
where $T$ is the plasma temperature, $\nu$ is the observing frequency, and EM is the emission measure \citep{Mezger1967}.  As the spectrum of the diffuse synchrotron emission is not significantly absorbed ($\tau<1$) at a frequency down to 330~MHz, we obtain EM$<2.9$ at $r\simeq100~{\rm pc}$ with $T=10^4$~K.  On the other hand, component NN peaks at $\sim660$~MHz in its spectrum \citep{Silver1998}.  Thus, we obtain EM$=12.7$ at $r_{d}=35$~pc, where $r_{d}$ is the projected distance of component NN from the core.

\subsubsection{Vertical Stratification}
Assuming that the electron density of the FFA plasma is uniform, we can ascribe the difference in EM to the difference in line-of-sight path length of the FFA plasma by a factor of $\gtrsim4$.  This can be realized if the plasma is extended spatially along the vertical direction and is shaped like a toroidal structure (Figure \ref{fig:Schematic}(a)). 

\cite{Izumi2018} recently showed that the CND of Circinus galaxy consists of the multiphase gas.  They found that the diffuse atomic carbon (CI) is spatially extended more in the vertical direction of the disk than the dense molecular gas probed by the CO(3-2) emission, which is consistent with the radiation-driven fountain model \citep{Wada2012}.  For NGC~1275, the atomic phase gas in the high CND latitude may be ionized by the ultraviolet (UV) irradiation from young and luminous star clusters around the AGN \citep{Richer1993}, which  then causes the FFA.  Or, the FFA plasma might originate in the ionized gas produced at further inner part of the CND/accretion disk and be outflowed by the AGN radiation.  There is evidence for such an ionized layer, with the lack of molecular gas, in the numerical simulations \citep[Fig. 1(c) in][]{Wada2016}.  The ionized outflows are indeed observed in the inner 500~pc of NGC~1275 by recent Gemini/NIFS observations \citep{Riffel2020}.  On the other hand, the relativistic electrons supplied by SN remnants would accumulate in the mid-plane of the disk because the star formation activity in the CND is likely more efficient in the dense gas region.  

\subsubsection{Density Gradient in the CND}
\begin{figure*}
\begin{center}
\includegraphics[width=17cm]{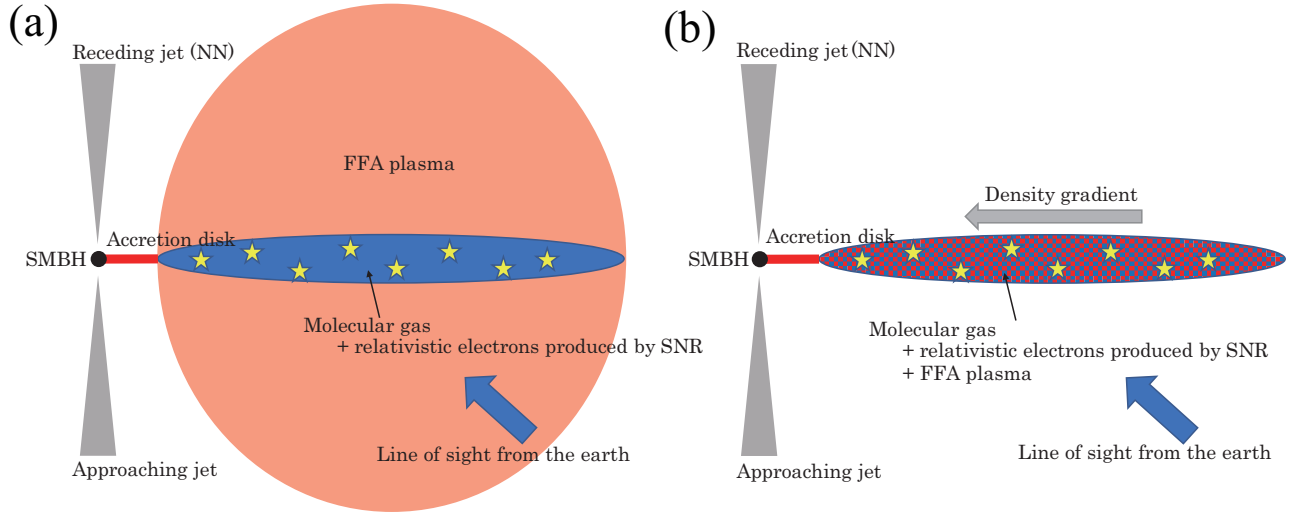}
\end{center}
\caption{Schematic picture of possible CND structure.  (a) The FFA plasma is more extended along the vertical direction than the molecular CND.  The scale radius of the FFA plasma structure is not necessarily to be the same, but it must be larger than $\sim30$~pc so that the plasma intercepts our line of sight to the receding jet (component NN in Fig. \ref{fig:fig1}). (b) The FFA plasma and molecular gas coexist  within the same CND plane with density increase toward the SMBH.}
\label{fig:Schematic}
\end{figure*}

Alternatively, the observed difference in EM can be explained as if the molecular gas, nonthermal electrons, and FFA plasma are mixed in the same CND plane and their density becomes higher toward the inner radius of the CND (Figure \ref{fig:Schematic}(b)).  In this case, FFA plasma is likely produced by the UV radiation from the stars in the CND.

We cannot discriminate between the two scenarios mentioned above with the present data.  A spectral study of diffuse synchrotron emission with more resolution elements and greater sensitivity would help to improve our understanding of the distribution of the FFA matter in relation to the CND.  Future high dynamic range observations with VLBI at $\lesssim1$~GHz with the Square Killometer Array (SKA) and next-generation Very Large Array (ngVLA) or similar study of other AGNs would be useful for unveiling a more detailed picture of CNDs.

\bigskip 
This paper makes use of the ALMA data of ADS/JAO.ALMA\#2017.0.01257.S. ALMA is a partnership of ESO (representing its member states), NSF (USA) and NINS (Japan), together with NRC (Canada), MOST and ASIAA (Taiwan), and KASI (Republic of Korea), in cooperation with the Republic of Chile. The Joint ALMA Observatory is operated by ESO, AUI/NRAO and NAOJ. This paper makes use of the VLBA data.  The National Radio Astronomy Observatory is a facility of the National Science Foundation operated under cooperative agreement by Associated Universities, Inc.  HN is supported by JSPS KAKENHI Grant Number JP18K03709.  NK is supported by JSPS KAKENHI Grant Number JP19K03918.

\renewcommand{\bibname}{}

\end{document}